# HOLOGRAPHIC DIGITAL FOURIER MICROSCOPY FOR SELECTIVE IMAGING OF BIOLOGICAL TISSUE

Sergey A. Alexandrov*, P. Meredith, T. J. McIntyre, and A. V. Zvyagin

Centre for Biophotonics and Laser Science, School of Physical Sciences, The University of Queensland,

Brisbane, QLD 4072

* School of Electrical, Electronic and Computer Engineering, The University of Western Australia, 35 Stirling Highway, Crawley, WA 6009

Phone: +61-7-3365 3430,  Email:  zvyagin@physics.uq.edu.au

**Abstract**

This paper presents an application of digital Fourier holography for selective imaging of scatterers with different sizes in turbid media such as biological tissues. A combination of Fourier holography and high-resolution digital recording, digital Fourier microscopy (DFM) permits crucial flexibility in applying filtering to highlight scatterers of interest in the tissue. The high-resolution digital hologram is a result of the collation of Fourier holographic frames to form a large-size composite hologram. It is expected that DFM has an improved signal-to-noise ratio as compared to conventional direct digital imaging, e.g. phase microscopy, as applied to imaging of small-size objects. The demonstration of the Fourier filtering capacity of DFM using a biological phantom represents the main focus of this paper.

# 1. Introduction

Imaging of biological tissue is a key research direction in biophotonics, wherein the main challenge arises from the fact that biological tissues are usually turbid. Biological tissue contains a vast range of scatterers of various shapes and sizes. Depending upon the effective refractive index of the constituent scatterers, a biological specimen is perceived as being either turbid or translucent. In turbid media, the depth of imaging is limited. The only plausible strategy to image through turbidity is to employ a mechanism that discriminates (gates) a selected elementary volume from within the obscuring environment of the medium.

There exist several implementations of this gating strategy. Among these, coherence gating is implemented in a recently emerged technique, optical coherence tomography, and has resulted in the production of clear images of such turbid tissues as human skin at sub-millimeter depths *in vivo* (Huang et al., 1991). The confocal gate is implemented in confocal microscopy, a technique that has been well established and instrumental in the production of *in vivo* images of superb sub-micron resolution (Langley et al., 2001). Labelling followed by microscopic imaging represents another approach widely used in modern biology (Zhang et al., 2002). Each of these approaches has limitations that determine their application niches. Despite the recent achievement (Povazay et al., 2002) attainment of micron-scale axial resolution remains technically difficult in optical coherence tomography. Confocal microscopy is complex, costly, and relies on a high-power light source. In addition, both of these techniques' gating mechanisms prove inefficient when applied to translucent specimens that provide poor imaging contrast. Labelling is time-consuming, invasive and introduces artifacts.

Most recently, a novel Fourier gating approach has been demonstrated by Boustany *et al.*, termed optical scatter imaging (OSI) (Boustany et al., 2001). The wide-angle and narrow-angle optical scatter contributions from a cell culture specimen were selected using correspondingly large and small aperture diaphragms in the Fourier plane of the objective lens. It has been shown that the ratio of these angular scatters is sensitive to minute size variations of the specimen constituents. Using this technique, swelling of sub-micron cell organelles that occurred after cell apoptosis has been detected. The power of OSI is derived from its capability to manipulate the resultant image in the optical Fourier domain. It can provide additional information on the specimen constituents and, ultimately, lead to enhanced imaging (Boustany et al., 2001, Valentine et al., 2001). However, this technique has several drawbacks. It relies on two sequential image acquisitions at each diaphragm opening, which is slow in its current implementation, or difficult to perform rapidly. Also, the acquired image pair optimally targets only scatterers of a certain size and shape. Selective imaging of non-circular scatterers remains a challenge using this technique. Generally speaking, the current implementation of Fourier gating has limited flexibility.

In this paper, we introduce a new implementation of the Fourier gating approach based on application of digital Fourier holography. Principles, methods and algorithms of digital holography have been described by (Yaroslavsky, 2003). Application of digital holography for measurements of large planar objects has been demonstrated by (Pedrini et al., 1999). The possibilities of digital holographic microscopy for imaging of micro objects with resolution equal to the diffraction limit of the imaging system have been also demonstrated (Cuche et al., 1999). However, it is for the first time to our best knowledge, the application of the digital Fourier holography for optical scatter imaging is addressed in this paper. The novel approach, termed digital Fourier microscopy, is particularly suitable for selective imaging of structures of various form-factors (size and shape) in biological tissues.

A hologram of the biological specimen is recorded using a digital recording device, e.g. a CCD camera, placed in the Fourier plane, i.e. back focal plane, of an objective lens that contains the specimen in its front focal plane, as shown in Figure 1. The reference wave is a plane wave incident on the recording device at an oblique angle (Fig. 1). It is also possible to record several collateral frames of the optical field in the Fourier plane to form a large-scale, high-resolution hologram. An image of the specimen is formed as a result of the reconstruction of the recorded hologram in software following digital filtering of targeted scatterers.

The application of holography represents a major departure from previous Fourier gating approaches (Boustany et al., 2001, Valentine et al., 2001) and confers several important advantages. Most notably, the digital format of the hologram recording permits crucial flexibility in applying filtering, which can be performed at the post processing stage followed by the reconstruction of the recorded image of a specimen (Goodman, 1996, Bovik and Acton, 2000). Therefore, the problem of the OSI limited flexibility is solved by employing powerful digital filters, which can be designed to target scatterers of interest. In contrast to OSI, the filter operation and the hologram acquisition are decoupled, which enables the multiple application of filtering to target scatterers of interest optimally. Secondly, DFM is expected to have an improved signal-to-noise ratio (SNR) in comparison with conventional direct digital imaging (DDI), e.g. phase microscopy, as applied to imaging of small-size scatterers. Therefore, DFM is potentially more sensitive and versatile than the Fourier gating approaches reported previously.

The aim of this paper is twofold: the introduction of digital Fourier microscopy, discussion of its merits, limitations and possible applications and; an experimental demonstration of the DFM Fourier gating capacity.

## 2. Digital Fourier microscopy

A biological specimen is placed at the object plane of a lens $L_F$, of focal length $f$, and illuminated by a plane wave whose wavevector is directed along the optical axis (see Fig. 1). A fraction of the illumination wave is split using a beamsplitter BS, and coupled into an optical fiber OF, to be conditioned as the reference wave. The fiber's distal end is positioned at the object plane. It is convenient to describe the lateral position of this fiber reference point-source by a two-dimensional coordinate vector $\vec{x}_0$ with respect to the origin at $O_x$ (Fig. 1). Light emerging from this fiber is collimated on its pass through $L_F$ with a wavevector directed at an angle with respect to the optical axis. Light scattered by the specimen interferes with the reference wave at a digital sensor placed in the Fourier plane of $L_F$. The resultant interferometric pattern is denoted by $I(\vec{r})$, and in the particular case of Fourier holography takes the following simple form (Caulfield, 1979):

$$I_F(\vec{r}) \propto C + F^{(2)}\left(\frac{\vec{r}}{\lambda f}\right)\cos\left(\frac{2\pi}{\lambda f}\vec{r}\cdot\vec{x}_0\right), \qquad (1)$$

where $\vec{r}$ denotes a two-dimensional coordinate vector in the digital sensor's plane, $C$ denotes the constant incoherent background devoid of information content. $F^{(2)}$

symbolizes a two-dimensional (2D) spatial Fourier transform, and $l$ denotes the illumination wavelength. A Fourier hologram is formed by recording of the intensity distribution given by Eq. (1). The unwanted contribution of the non-scattered illumination wave is eliminated via a filed stop in front of the sensor.

Firstly, analysis of Eq. (1) shows that DFM has the inherent capability of subsurface layer gating. A constant factor under the exponential sign, $\mathbf{n} = |\vec{x}_0|/lf$, termed the hologram spatial frequency, accounts for this property. If we assume for the moment that the object is a point-like scatterer, Eq. (1) describes the hologram that represents a family of equally spaced linear fringes (spatial tone of frequency $\mathbf{n}$). This hologram is characterized by linear spatial phase distribution in the Fourier plane. At the same time, the hologram of an off-plane scatterer represents a family of curved interference fringes crowded together on the hologram plane as they move away from the optical axis. This hologram is characterized by a non-linear spatial phase distribution in the Fourier plane. Therefore, the contribution from the off-plane scatterers can be filtered out by applying a 2D spatial filter (Bovik and Acton, 2000). Alternatively, the digital filter can be designed to gate selected off-plane scatterers. By adjusting the numerical parameters of the 2D spatial filter, a three-dimensional tomogram of the specimen can be built from the single Fourier hologram.

DFM provides a means to realize Fourier gating, i.e. gating of scatterers of certain shape and size, by exploiting the digital format of the hologram. Eq. (1) shows that the 2D spatial Fourier transform factor, $F^{(2)}$ describes the amplitude modulation of the single spatial tone of the hologram. $F^{(2)}$ is determined by the scatterer size and shape. In the realistic case of a specimen comprising several scatterers, $F^{(2)}$ is a product of each scatterer's 2D Fourier transform, each of which is centered with respect to the optical axis. Application of the 2D mask calculated for a scatterer of interest, $F_0^{(2)}$, to the hologram, followed by subsequent hologram reconstruction in software, will pick up only scatterers whose Fourier transform is conformal with $F_0^{(2)}$. To illustrate the significance of this capability, consider the recording of the Fourier hologram of a cell culture conventionally modelled as a random set of nuclei (spheroids) embedded in a lower refractive index cytoplasm. It is expected that the described digital processing followed by digital image reconstruction should pick up spheroids that match the pre-selected parameters of the digital mask, with the most important parameter being the spheroid size. The application of a set of digital masks calculated for a range of spheroid sizes yields the size distribution of the cell nuclei, which serves as the key diagnostic indicator of, for example, cancer (Backman et al., 2000).

The DFM permits 'framable' holographic recording, i.e. recording of a high-resolution, large-format, hologram comprising multiple frames captured by a digital camera at the collateral positions on the Fourier plane. The unique spatial structure of a Fourier hologram that is modulated by a single spatial tone, allows accurate computer-aided collation of the collateral frames. The size of the resultant composite hologram is virtually unlimited and only practically constrained by the optics aperture. Importantly, the multi-frame recording method circumvents the long-standing shortfall of low-resolution in digital holography associated with the insufficient sampling capacity of modern digital sensors (Pedrini et al., 1999).

The proposed DFM is expected to have an improved SNR in comparison with direct imaging modalities, e.g. phase microscopy, as applied to imaging of small size objects. The improvement results from the DFM capacity to detect a greater amount of power diffracted from a small-size object, as this power can be distributed over a greater number of pixels (see Fig. 2). In the case of a single-pixel object, DFM can ultimately accommodate a factor of $MN$ more photoelectrons on the digital $M \times N$ sensor than the DDI configuration before it is saturated. Here, we assume that a one-pixel object is optical-Fourier-transformed into a constant-amplitude signal over the entire area of the sensor, and is modulated by a single spatial tone. In the DDI configuration, the illumination intensity must be reduced to prevent saturation of the signal. Under the reasonable assumption that the reference wave intensity is controlled to be the same for both configurations, the following expressions for the number of photoelectrons in the DDI and DFM configurations, respectively, can be written:

$$\boldsymbol{f}_{DDI} = \begin{cases} \boldsymbol{h}\left(I_0 + \dfrac{\boldsymbol{s}}{S_{px}}I_0 + 2I_0\sqrt{\dfrac{\boldsymbol{s}}{S_{px}}}\right), & \text{at one pixel} \\ \boldsymbol{h}I_0, & \text{otherwise} \end{cases}, \qquad (2)$$

$$\boldsymbol{f}_{DFM} = \boldsymbol{h}\left(I_0 + \dfrac{\boldsymbol{s}}{S_{px}MN}I_1 + 2\sqrt{\dfrac{I_0 I_1}{S_{px}MN}}\cos\left[\dfrac{2\boldsymbol{p}}{\boldsymbol{l}f}\vec{r}\cdot\vec{\boldsymbol{x}}_0\right]\right), \qquad (3)$$

where $I$ stands for light field intensity: $I_0$, $I_1$ denote the reference field and DFM illumination intensities, respectively. $\boldsymbol{s}$ denotes the differential scattering cross-section integrated over the detection optics acceptance angles. $\boldsymbol{h}$ is the conversion factor that depends on the quantum efficiency of the sensor, exposure time, and pixel area. $S_{px}$ is the area of a single pixel. In deriving Eq. (2), we assumed that the illumination and diffraction optical waves interfere constructively. The 3$^{rd}$ term on the right hand side of Eq. (3) represents the DFM signal. The digital Fourier transform of this signal term results in the reconstruction of the one-pixel object, which takes the form of (Bracewell, 1978)

$$\boldsymbol{f}_{DFM,F} = 2\boldsymbol{h}\sqrt{\frac{I_0 I_1 \boldsymbol{s}\, MN}{S_{px}}}\,. \qquad (4)$$

Shot-noise is assumed to be the dominant noise in the detection circuit. It originates from the incoherent background due to the reference optical field. The shot-noise is expressed by the first term of the right-hand side in Eqs. (3) and (4), and is equal for both configurations. The digital Fourier transform of the shot-noise results in

$$\langle \boldsymbol{s} \rangle^2_{shot,F} = \boldsymbol{h} I_0 MN\,. \qquad (5)$$

Based on the analysis presented above, comparison of the signal-to-noise ratios for DFM and DDI configurations yields the following estimate:

$$\frac{SNR_{DFM}}{SNR_{DDI}} = \frac{I_1}{4 I_0} \qquad (6)$$

The factor of 4 in the denominator of Eq. (6) is because the DFM-reconstructed image occupies only a quadrant of the DDI image. Eq. (6) is obtained using several simplifying assumptions, and will vary widely depending on experimental conditions. Using typical experimental parameters, as detailed in the experimental part of this paper, DFM has a potential SNR improvement factor of 10-20 dB over the DDI configuration, where the typical saturation optical energy level of a 12-bit CCD camera (Dalsa) was assumed. It is interesting to note that DFM represents a realization of the optimal SNR detection strategy in the spatial/frequency domain, whereas in the time/frequency domain, the superior SNR performance of the Fourier-domain OCT has been recently demonstrated (Leitgeb et al., 2003, de Boer et al., 2003).

## 3. Experiment

We commenced the experimental study of DFM by testing its Fourier gating property. The experimental setup is schematically shown in Fig. 3, and consists of three major parts: an optical circuit, a digital recording device, and a specimen.

The core part of the optical circuit is the Fourier optical configuration, as shown in Fig. 1. Analogously, a Fourier hologram of the magnified image of a specimen was formed on the recording device. An achromat doublet lens of 60-mm focal length was used as the Fourier lens, $L_F$ that contained the magnified image of the specimen and the recording device in its front and back focal plane, respectively. The magnified image of the specimen was formed on the image plane via a telescopic assembly comprising a lens pair: objective lens, $L_{obj}$ (oil-immersion, infinity-corrected, $f_{obj} = 3$ mm, $N.A. = 1.4$) and relay lens, $L_r$ (singlet lens, $f_r = 60$ mm). The main function of the assembly was to combine the specimen's image and reference point-source at the front focal plane of $L_F$, and to control lateral separation between them. The separation between $L_{obj}$ and $L_r$ was set to $f_{obj} + f_r$, which determines the magnification factor of $f_r/f_{obj} = 20$, where the specimen was situated at the front focal plane of $L_{obj}$. The specimen was illuminated by a collimated Gaussian laser beam (Helium-Neon laser, 10 mW) contracted via a reduction telescopic assembly (not shown in Fig. 3) to ~0.1-mm $1/e^2$ beam diameter. The illumination intensity was controlled by a neutral density filter set. A controlled amount of the laser beam intensity was split and coupled into a single-mode fiber. The fiber's distal end was situated off axis ($|\vec{x}_0| \approx 2\ mm$) at the image plane. The outgoing light beam from the fiber produced the reference beam, a collimated Gaussian light beam incident on the recording plane at an angle. The reference and object optical fields interfered at the recording device producing a holographic pattern.

To capture the hologram we used a two-step procedure (in place of a high-dynamic-range high-resolution CCD camera that was not available): Firstly, an analogue hologram was recorded on a conventional 50×50-mm photosensitive plate (PFG-01, Slavich). We controlled the optical field intensity and exposure to ensure the optimal dynamic range for the recording. The exposure was realized by employing a timer-controlled mechanical shutter operated on the sub-second scale. The exposed plate was subsequently developed. As a next step, the recorded hologram was digitized by using a scanning transmission microscope to acquire a series of 12-bit frames of 1024×1024 size, which were subsequently stored in a personal computer. The recorded hologram featured the spatial fringe period of about 20 μm. Each fringe was digitally sampled with 8 pixels. In order to realize a multi-frame digital hologram, we acquired 35 collateral frames and merged them together in a 2290×3122-size composite hologram using computer code written in Matlab. The core function of this code was co-location of the overlapping fragments of the adjacent frames by employing the cross-correlation procedure. In order to match the uneven brightness of the acquired frames, the brightness level of each frame was adjusted using a 2D Gaussian apodization function.

The specimen was a cell culture phantom created by dispersing a mixture of 5.26 μm and 2.01μm polystyrene spheres in polyvinylalcohol (PVA) and depositing the resulting composite by spin coating onto quartz cover slips to produce solid thin films. As such, the phantom mimicked a cell nuclei suspension in a cytoplasm matrix, and can be considered as a realistic representation for moderate resolution optical imaging purposes. The spheres were obtained as a 5% weight / volume colloidal dispersion (latex) in water (ProSciTech, Australia). The fabrication procedure was as follows: A 10% by weight solution of PVA (22,000 MW analytical grade) was made by dissolving the polymer in de-ionized water (18.2 MΩ) under vigorous stirring at 90°C for 2 hours. Once a clear solution had been achieved, the PVA dispersion was allowed to cool to room temperature and the latex added in a volume ratio of 50:1:1 (PVA : 5.26 μm latex: 2.01 μm latex). This yielded a final fractional weight concentration of both spheres in the phantom of $1 \times 10^{-5}$. Thin films were cast by spinning the cooled solution (a milky colour) onto cleaned quartz cover slips.

PVA films containing no latex were also fabricated in the above manner in order to calculate film thickness and refractive index. This was done by measuring the optical transmission of the PVA film in the wavelength range 450 nm to 900 nm using a Perkin Elmer λ40 spectrophotometer. The thin film optical constants were extracted from these transmission measurements using the turning point analysis of (Meredith et al, 1993). It was found that the films deposited under these conditions had an average refractive index of 1.515±0.005 at $\lambda$ = 633 nm and were ~1.1 ± 0.1 μm thick.

## 4. Results and Discussion

In order to reconstruct the image of a specimen it was sufficient to perform a 2D Fourier transformation in software. A fragment of the reconstructed image is presented in Fig. 4 (a). Magnified images of one large (5.26 µm) and three small (2.01 µm) spheres are clearly resolvable (indicated by arrows). Immediately prior to the hologram capture, the image layout was viewed directly to ensure its identity with the reconstructed layout. The direct viewing was realized by projecting the magnified image of the specimen onto a remote screen with $L_F$ and the holographic plate removed (see Fig. 3). In order to apply 2D Fourier filtering, we designed digital masks based on Mie-scattering calculations. The small-sphere mask was designed by calculating the phase scattering function of a 2.01-µm polystyrene sphere of refractive index of $1.59 \pm 0.01$ at $\lambda = 633$ nm embedded into the PVA host matrix. To calculate a digital mask for the large sphere, we assumed the host medium was air of refractive index unity. Results of the Mie-calculations are presented in Fig. 5. The phase scattering functions of a 2.01-µm sphere (dashed line) and 5.26-µm sphere (solid line) are plotted versus the scattering angle. The scattering functions are normalized by equating their integrals over the full solid angle to unity. In order to assess the efficiency of the calculated digital filters, we performed a comparison with the angular scatter of the spheres obtained experimentally using the following method: In the reconstructed digital image, the sphere of interest was cropped. An inverse 2D Fourier transform of the cropped image yielded the selected sphere contribution to the composite Fourier hologram. The spatial frequency modulation was removed from the selective holographic transform by sequential application of a spatial band-pass filter, square-law detector, and low-pass filter. The inset of Fig. 5 shows the resultant angular scatter intensity map of the small and large spheres. The two maps are montaged together to facilitate their comparative viewing. Note their distinctly different angular scatter patterns that underpins the anticipated efficiency of the Fourier gating operation, and in accord with the OSI technique (Boustany et al., 2001). The scattering angles corresponding to the peaks and troughs of the small and large spheres angular scatters (characteristic scattering angles) are marked by arrows in Fig. 5. The characteristic scattering angles of the theoretical curves are labeled. In order to facilitate comparison between the theoretical and experimental results, the arrows representing the characteristic experimental data are correspondingly labeled. The comparison shows reasonable agreement between the calculated and experimental data. The discrepancy is noticeable at the high scattering angles, and is attributed to the following reason. We considered the spheres embedded in the homogeneous infinite medium in the Mie theory framework, whereas in our case, the small and large spheres were deposited on the cover slip and embedded in a thin layer of 1-$\mu$m thickness. Therefore, the results of the Mie-scattering calculation need to be corrected for mode coupling between the spheres, finite-size PVA layer, and glass surface (Bass et al., 2003, Jory et al., 2001).

In order to demonstrate the Fourier gating we, first, calculated the filtered Fourier transform, which is a product of the calculated 2.01-$\mu$m sphere scattering phase function and the acquired composite hologram. The filtered image is an inverse 2D Fourier transform. In the direct image representation, the filtered image is a convolution of the direct calculated image of the sphere and the reconstructed direct image. (Rigorously speaking, one has to calculate the Mie scattering amplitudes and compute their Fourier-transforms in order to obtain the optical Fourier transform of the sphere.) At the second step, we compute the ratio of the original reconstructed image and the filtered image, so that the digital pick up image is obtained [c.f. the ratio of the two angular scatters, as described by (Boustany et al., 2001)]. Fig. 4 (b) presents an example of this digital pick up image, which shows complete suppression of the contribution from the small spheres, whereas an image of the large sphere remains clearly resolvable (white arrow). We have also verified that the similar gating performance could be achieved by using a 2.01-µm digital mask extracted from the experimental data, instead of the calculated digital mask. It demonstrates that the DFM technique has flexibility in the context of the filter design. One can design a spatial filter using prior knowledge about the specimen, e.g. targeting scatterers of interest. Alternatively, the digital filter can be designed using a template specimen to pick up the difference between the template specimen and the specimen under study.

Several problems remain to be solved before DFM can be used for biomedical applications. The superior SNR in comparison with direct imaging modalities needs to be demonstrated. In the present experimental study, the SNR of the reconstructed image was poor due to the following reasons. We used a two-step recording process that resulted in excessive noise of the composite hologram. The dynamic range of our data (12-bit) was not sufficient to record small signals in the presence of a large background. This difficulty was exacerbated by the fact that the reference optical field was Gaussian of ~15 mm $1/e^2$ diameter, and, therefore, exhibited an unwanted intensity gradient from the center outwards. It was difficult to account for this intensity gradient when merging the acquired frames.

For an accurate quantitative study of the DFM performance, the cell-culture phantom has to be improved. A thicker layer of PVA in which both types of spheres are completely immersed should result in a more accurate agreement of the experimental scatter data and Mie-scattering-based theoretical calculations; and, secondly, a more sophisticated theoretical model that takes into account surface mode coupling has to be implemented into the theoretical analysis (Bass et al., 2003, Jory et al., 2001).

## 5. Conclusion

We have proposed and demonstrated an application of digital holographic Fourier microscopy, for selective imaging of scatterers in biological tissue. This technique, DFM, can be classified as a Fourier gating technique. Among the technique's merits, improved signal-to-noise ratio, sub-surface gating capacity, and high-resolution recording based on multi-frame collation, makes it a promising imaging modality for biomedical implementation. In this paper, we have demonstrated the Fourier gating capability using a biological phantom. Our experimental results suggest that the demonstrated gating capacity can be useful in the study of selective imaging of the nuclei size distributions in cell cultures with a particular goal to aid histology of cancer.

FIGURE CAPTIONS

**Figure 1.** Schematic diagram of a DFM microscope. $\xi$, r in combination with the dotted lines mark frames of references associated with object and digital sensor, respectively. OF – optical fiber; BS – beamsplitter.

**Figure 2.** Schematic diagrams of (a) direct imaging and (b) DFM modalities for comparison of their corresponding signal-to-noise ratios.

**Figure 3.** Experimental setup and ray diagram of the DFM microscope. Illumination and marginal rays are shown by grey and black arrow lines.

**Figure 4.** (a) Reconstructed holographic image of the cell culture phantom: 5.26-$\mu$m and 2.01-$\mu$m polystyrene spheres (arrows) in PVA matrix; (b) Digital pickup image showing that only the large sphere (arrow) remains.

**Figure 5**. Plot of the calculated scattering phase functions of 2.01- and 5.26-$\mu$m size spheres versus scattering angle. Arrows indicate the angular coordinates of the maxima and minima determined from the experimental data.
Inset (top right corner): digital masks extracted from the experimental data. Left-hand semicircle, 2.01-$\mu$m sphere; right-hand semicircle, 5.26-$\mu$m sphere, referred respectively as 2-$\mu$m- and 5-$\mu$m- sphere.

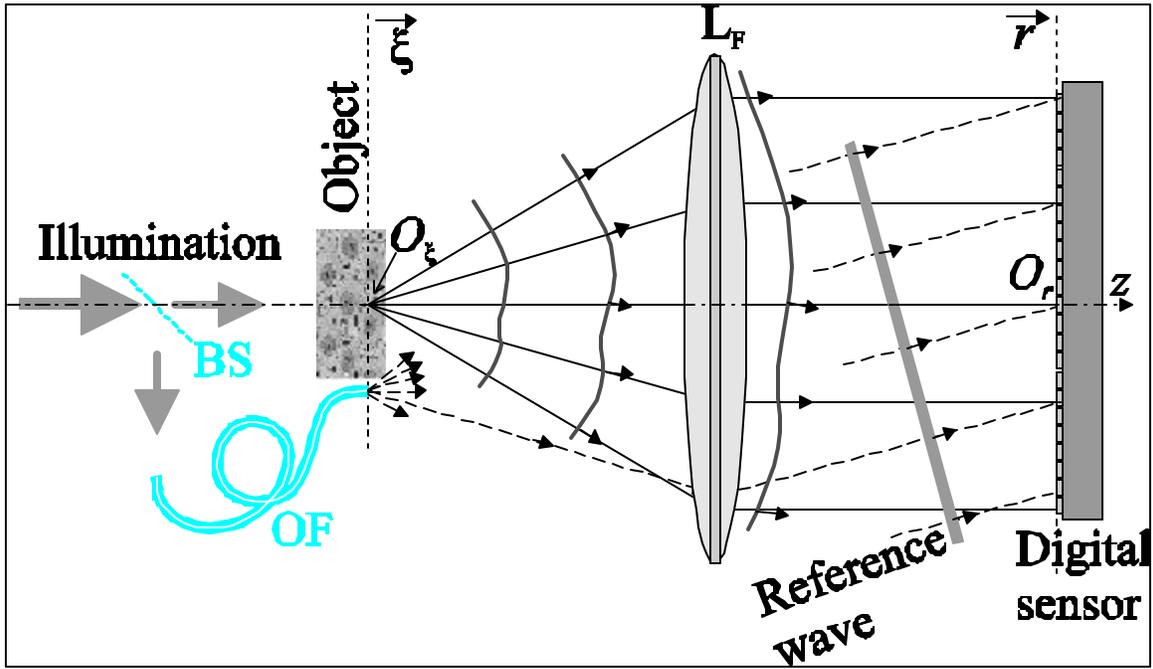

Figure 1.

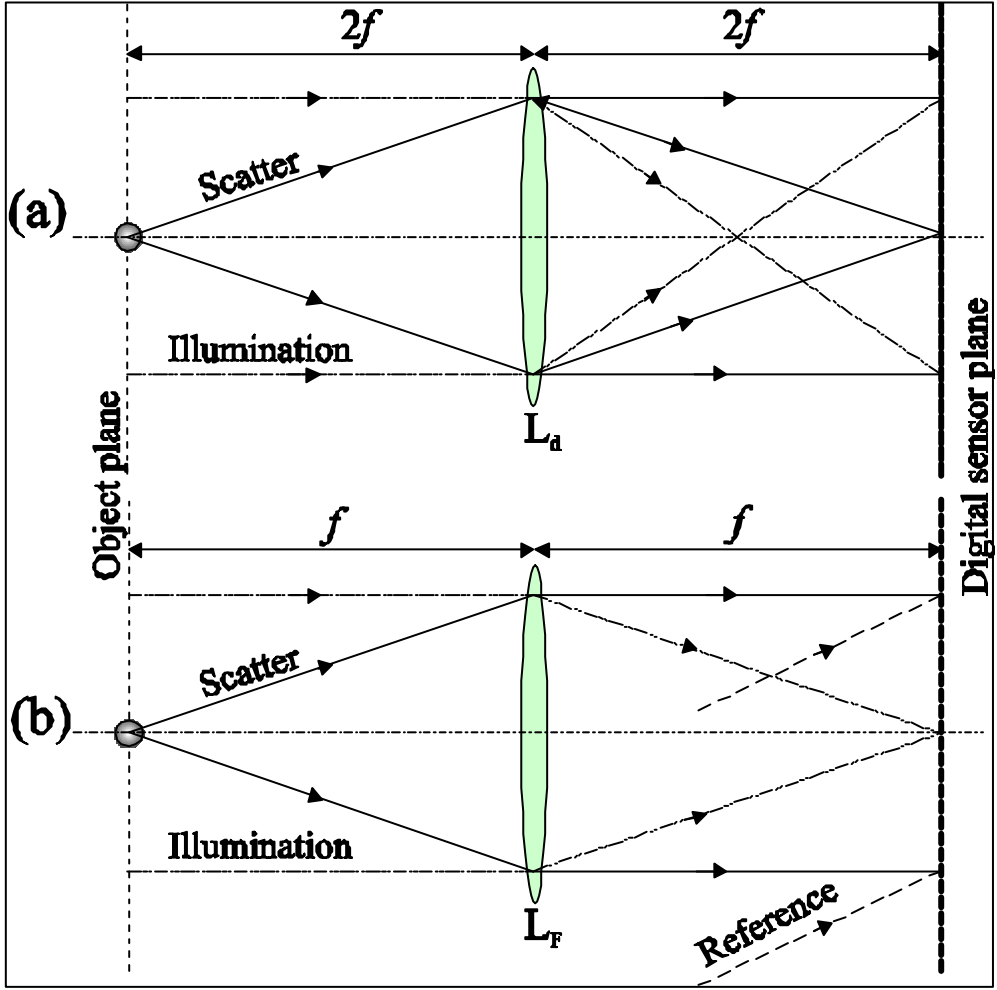

Figure 2.

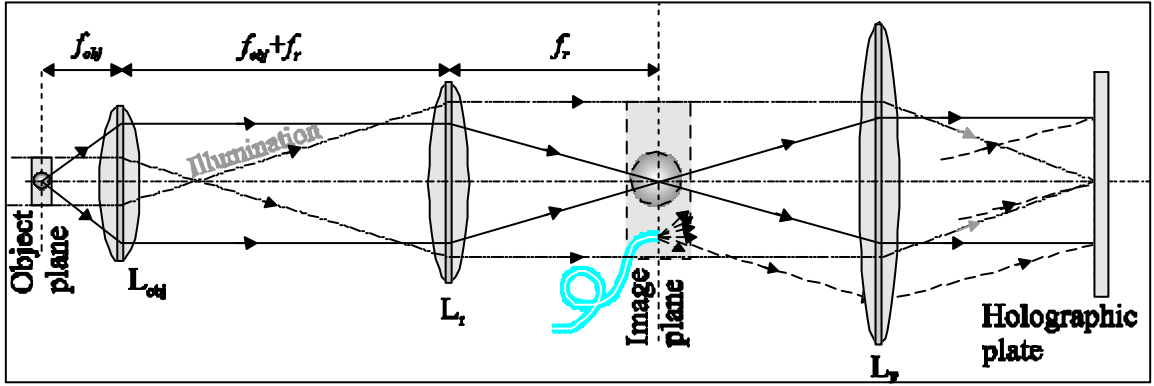

Figure 3.

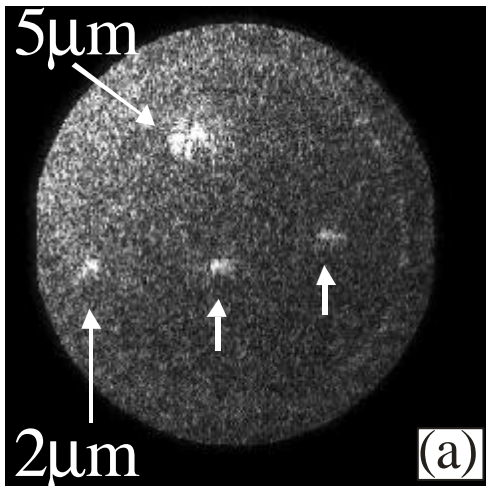 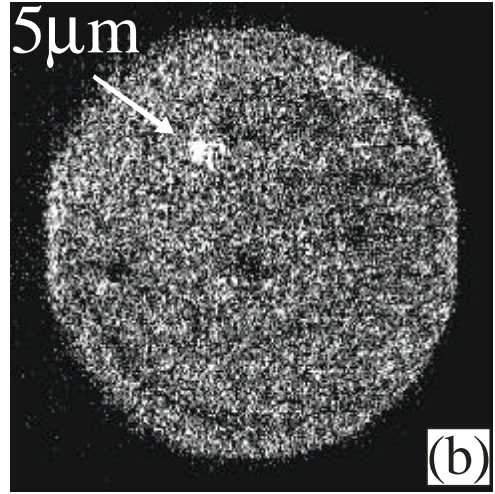

Figure 4.

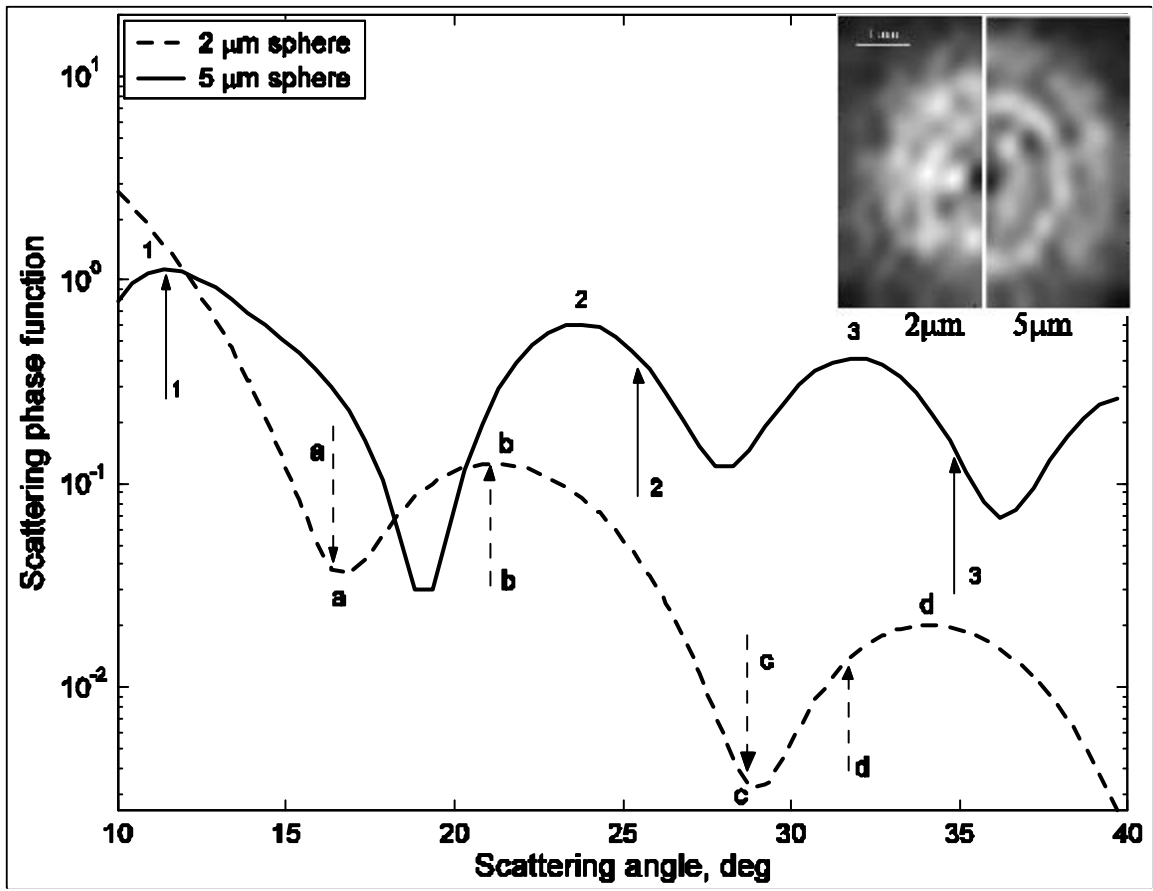

Figure 5.